\documentclass[aps,prb,showpacs,twocolumn,superscriptaddress]{revtex4-1}
\usepackage{bm,color,amsmath,amssymb,mathrsfs,latexsym,graphicx,psfrag,mathtools}
\usepackage{graphicx}
\usepackage{bm}
\usepackage{ulem}
\usepackage{color}
\usepackage{amsmath}
\usepackage{epsfig}
\usepackage{epstopdf}
\usepackage{comment}
\newcommand{\bea}{\begin{eqnarray}}
\newcommand{\eea}{\end{eqnarray}}

\begin{document}

\title{Superfluid Density, Penetration Depth, Condensate Density}
\author{Warren E. Pickett }
\affiliation{Department of Physics and Astronomy, 
University of California Davis, Davis CA 95616}

\begin{abstract}
Fascination with the concept of superconducting 
(SC) {\it superfluid density} $\rho_s$ has persisted
since the beginning of superconductivity theory, with numerical values
of an actual density rarely provided. Over time $\rho_s$, addressed mostly
in cuprate and following high temperature superconductors, has become
synonymous with the normalized (unitless) inverse square of the magnetic
penetration depth $\lambda_L$ (the London expression, with superfluid
density denoted $n_s$), 
with interest primarily on its temperature $T$ dependence that is
expected to reflect the T-dependence of the SC gap 
amplitude and gap symmetry. 
In conventional superconductors, generalized expressions from the London
penetration depth via Ginzburg-Landau theory, then to BCS theory provide
updated pictures of the supercurrent density--vector potential relationship. 
The BCS value $\lambda_{band}$ is distinct from any particle 
density, instead involving particle availability at the Fermi surface and Fermi velocity 
as the determining factors, thus providing a basis for a more fundamental
theory and understanding of what is being probed in penetration depth studies.
The number density of superconducting electrons ${\cal N}_s(T$=0) 
-- the scalar SC {\it condensate density} --
is provided, first from a phenomenological estimate but then supported 
by BCS theory. A straightforward relation connecting ${\cal N}_s(0)$ 
to the density of dynamically transporting carriers
in the normal state at $T_c$ is obtained.
Numerical values of relevant material parameters including 
$\lambda_{band}$ and ${\cal N}_s$ are provided for a few conventional SCs.
\end{abstract}

\date{\today}
\maketitle

%\newpage

\section{Introduction}
\label{sec:intro}
This paper compares and contrasts three properties of
the title that involve the ``superfluid stuff" of a superconductor: 
superfluid density, of great interest but without a microscopic
definition; penetration depth, the subject of various probes but 
challenging to measure directly; condensate density, a quantity
of little previous treatment or interest.
The introduction of magnetostatics into description of the
superconducting (SC) state by the London brothers in 1935 
proposed\cite{London1935} that the supercurrent
density $\vec J_s(\vec r)$  is directly proportional to a properly 
gauged vector potential $\vec A(\vec r)$,
connected by a magnetic field penetration depth $\lambda$, given
by them as the time as $c^2/\lambda^2=4\pi e^2 n_s/m_s$, with superconducting density
$n_s$ and mass $m_s$ arising from the atomistic picture of electrons. 

In 1934 Gorter and Casimir (GC) had introduced\cite{Gorter1934}
the concept of a separation of the total density of charge ``carriers''
(without specification) in the superconducting state $n$=$n_s+n_n$, 
the SC and normal components, each
contributing differently to thermodynamics. 
They introduced the phenomenological GC superfluid fraction $x=n_s/n$
and arrived at a free energy expression in those terms, according to
their understanding at the time. Schrieffer noted\cite{Schrieffer1964}
that BCS theory provides different expressions for the SC state free
energy than used by GC, both through different dependencies on $x$
and through more modern free energy expressions. 

In spite of the continuous development of electronic 
structure theory and the sporadic extensions of SC theory and 
formalism (see Sec. II), an absolute superfluid density $n_s$ and the associated
mass $m_s$ have remained concepts of fascination but with little quantification,
while the penetration depth gained acceptance as a tensor quantity.
Brandt has given an
example, and improved analysis,\cite{Brandt1988} of theory that leads to the 
 determination of the penetration
depth from muon relaxation spectroscopy without any mention of 
the London parameters, based on the
observables of magnetic vortices and their lattices. 
Conceptual issues arise. Super{\it current} density $\vec J_s$, from which 
the London picture arose, is an observable
for which a scalar super{\it fluid} density might seem to be an essential 
factor, but $\vec J_s$ has been recognized to be a tensor quantity. 

There has been a transformation in the literature of the term
``superfluid density.'' Initially, it was understood as the number density of 
superconducting electrons in the static condensate, not otherwise defined.
In 1948 F. London repeated\cite{London1948} that $n^L$ and $m^L$ 
(notation here, for later clarity) were
undefined quantities, and introduced the modern language of Fermi surface quantities, 
as will be discussed in this paper. As further evidence of the understanding
of this topic, Tinkham stated in his book,\cite{Tinkham1996} $n^L$ and $m^L$ are
simply `phenomenological parameters' of the London theory.
Over decades this (scalar) `superfluid density' label has been converted
into a dimensionless quantity $\rho_s$ with the same name, but moving emphasis from
a physical density, recognized as being difficult to quantify, to the 
temperature dependence of $\lambda$, reflecting the amplitude and temperature
dependence of the gap. The concept is that this penetration
depth variation is what is being probed, but the designation `superfluid density' has
been retained. 
 
The literature on superfluid density, superfluid fraction, and 
a related ``superfluid stiffness'' can be confusing.
Two examples relative to $n_s$ are illustrative.
Emery and Kivelson, in their paper\cite{Emery1995} 
on the ``superfluid stiffness'' of
underdoped cuprates -- which incorporated the London $n_s^L$ as a central 
property in their equations -- gave no values of $n_s$ among their
tabulation of related properties of 14 SCs.
Basov and Timusk, in a review of superfluid properties of cuprate
SCs,\cite{Basov2005} contains `superfluid density' or
`superfluid response' 67 times, without quoting any numerical
value\cite{Basov2} for $n_s$ or a related quantity. 
%%%%%(They provide an updated Uemura plot [see below] 
%%%%%  that contains a $\rho_s$ axis.) 
These two influential papers provide a vivid picture of just how elusive
quantification of absolute $n_s$ has been, and remains. 
A few experimental estimates of $n_s$ will be presented in 
Sec. \ref{sec:penetration}. 

To extend the conversation and perhaps reduce confusion we revive 
the rarely invoked {\it condensate density ${\cal N}_s$} --
the density of superconducting electrons -- and contrast it with
superfluid density as now treated in analysis of experimental data. 
Intuitively, it would seem that supercurrent density
would involve a charge and density of carriers and their velocity. 
This simple picture implies that the density as well as the velocity 
might be axis-dependent in non-cubic crystals, as were inferred penetration depths. 
A few quantitative examples
of ${\cal N}_s$ and penetration depths based on
modern electronic structure theory will be provided.

This paper is organized as follows. After a historical perspective of theory
provided in Sec.~\ref{sec:history}, the Drude plasma frequency arising
in the theory is discussed in Sec.~\ref{sec:Drude}. 
Some reported estimated of penetration depth in relation to the London superfluid
density are noted in Sec.~\ref{sec:penetration}.
Relevant far-IR optical properties are
discussed in Sec.~\ref{sec:farIR}, followed by introduction of the 
little discussed condensate density in Sec.~\ref{sec:condensate}. Example
values of these and other quantities appearing in this paper are provided in Table I
for selected Fermi liquid superconductors. Due to the great 
interest in superfluid density in high temperature superconductors, a
recounting of some empirical correlations in this area are given in Sec.~\ref{sec:exotic}.
A Summary in Sec.~\ref{sec:summary} completes the paper.

\section{Supercurrent density, superfluid density}
\label{sec:history}
\subsection{London theory}
{\it London theory} addressed the  magnetic
field exclusion from the bulk SC -- the recent (at that time) Meissner-Ochsenfeld 
observation\cite{Meissner} of exclusion/expulsion by the SC state.
The key realization was that the SC state reacts differently from the normal
state: instead of electrons responding to the magnetic flux density
$\vec B=\nabla\times\vec A$, the connection was to the vector 
potential $\vec A$: the supercurrent 
density arises linearly from the vector potential. 
The London result was\cite{London1935,Schrieffer1964} (our notation for
later clarity)
\bea
\vec J_s^L(\vec r)&=&-\frac{e^2}{c} \frac{n^L}{m^L} \vec A(\vec r) \equiv
           -\frac{c}{4\pi} \frac{1}{\lambda_L^2} \vec A(\vec r) \nonumber \\
  \frac{c^2}{\lambda_L^2}&\equiv& 4\pi e^2 \frac{n^L}{m^L}.
\label{eqn:London}
\eea
This remarkably simple equation (subject to other electromagnetic restrictions) 
introduces the inverse squared magnetic ($B$) field penetration depth $\lambda_L^{-2}$ times
$\vec A$ (in the divergenceless gauge).
$\lambda_L$ was envisioned as a material dependent length characteristic
of the {\it SC state}. The
relation involves fundamental constants and the unspecified superfluid
density $n^L$ and mass $m^L$. 
The value of $n^L$ would be expected to decrease from maximum at
T=0 to vanishing at T$_c$. The Londons noted\cite{London1935} 
that ``there is no particular reason, for attributing to 
[$\lambda_L^{-2}$] the value $n/m$ [their notation]'' except for their 
atomistic picture at the time. 

A common model independent prescription for the penetration depth is
\bea
\frac{1}{\lambda(T)} = \int_0^{\infty} \frac{B(z,T)}{B}dz,
\label{eqn:formallambda}
\eea 
formulated for the textbook case of a uniform field $B$ applied to a
SC at $z>0$ with a smooth surface at $z$=0. This expression generalizes
the specification to non-exponential decay $B(z,T)$. 
This profile has been mostly inaccessible, so experimental identifications
have used indirect procedures to obtain estimates of $\lambda(T)$.  

As mentioned in the Introduction, use of ``superfluid density'' in the literature
has evolved from $n^L(T)$ to that of a normalized (unitless) quantity 
$\lambda^2_{exp}(0)/\lambda^2_{exp}(T) \equiv \rho_s$. With the
discovery of superconducting gaps with a T-dependence that differs
from that of the BCS gap $2\Delta(T)$ due to anisotropy or nodes, 
attention turned to $\rho_s(T)$ as a 
probe of exotic order parameters (energy gaps) to account for their
T-dependence.  $\rho_s$ remains almost universally 
discussed in terms of the London picture of Eq.~\ref{eqn:London},
and referred to as ``superfluid density.'' As a specific example,
``low superfluid density'' is used as a characterization of the
underdoped cuprate superconductors.

\subsection{Ginzburg-Landau penetration depth}
{\it Ginzburg-Landau theory} (GL)\cite{Ginzburg1948} was derived 
from a free energy functional of the
complex SC order parameter $\Psi(\vec r,T)$ and the vector potential $\vec A(\vec r)$.
The vector potential was chosen, as in classical electrodynamics and then quantum
mechanics, to appear as an addition to the momentum operator $-i\hbar\nabla$ in the
canonical momentum, coupling through particles of charge $e^*$ and mass $m^*$,
presciently anticipating that these might not be simple electrons.

Minimization of this functional with respect to both functions 
leads\cite{Ketterson1999} to the supercurrent density expression, involving the phase
$\eta(\vec r)$ of $\Psi$, as (with $T$ dependences suppressed)
\bea
\vec J_s^{GL}(\vec r)&=&\frac {e^*\hbar}{2m^*} Im[\Psi^*(\vec r)\nabla\Psi(\vec r)]
   -\frac {(e^*)^2}{c} \frac{|\Psi(\vec r)|^2}{m^*} \vec A(\vec r)\nonumber \\
 &=&e^*\frac{\big[\hbar\nabla\eta(\vec r)-\frac{e^*}{c}\vec A(\vec r)\big]}{m^*}
         |\Psi(\vec r)|^2.
%%\left[\frac{e\hbar}{2 m}|\Psi(\vec r)|^2 ~~\left(\nabla\eta(\vec r)
%%      -\frac{2e}{c}\vec A(\vec r)\right)\right],
\eea
Given the association of $|\Psi|^2$ with $n^L/2$ (see below), 
the last term gives rise to a London-like penetration depth proportionality
between $\vec J_s$ and $\vec A$, {\it except} for 
$\vec r$ and T dependence of $|\Psi(\vec r)|^2$ and $\eta$. 
The first term introduces the superfluid phase softness
(inverse of stiffness) $\nabla\eta(\vec r)$, also proportional to $\vec A$ 
 (or arising from boundary conditions which give position dependence to
$\Psi(\vec r)$).  The quantity in square brackets, when divided
by $m^*$, is the Cooper pair velocity, giving $J_s$, conceptually correctly,  as
charge $\times$ velocity $\times$ density.
This new term in $\nabla\eta$ provides a second contribution to the
generalized penetration depth linking $\vec J_s$ to $\vec A$, giving
$\lambda^{-2}_{GL}(\vec r,T)$.

\subsection{BCS relation between $\vec J_s$ and $\vec A$}
{\it BCS theory}\cite{BCS1957} is the basic microscopic quantum theory
of SC, based on a many-body wavefunction comprised of strongly correlated 
Cooper pairs and non-superconducting quasiparticles (QPs)
(broken Cooper pairs). With supercurrent already addressed 
partially by BCS, the relation to $\vec A$ 
was carried through fully by Chandrasekhar and Einzel\cite{Chandra1993} in terms of
the electronic spectrum of the normal state, BCS QP thermal occupation factors,
and the SC energy gap, for arbitrary Fermi surfaces.
For simplicity, band subscripts are not displayed in the following equations.

The derivation requires the QP dispersion relation $E_k$, with
$E_k^2=\varepsilon_k^2+\Delta_k^2$ in terms of the normal state
energy $\varepsilon_k$ relative to the Fermi energy and the energy
gap $2\Delta_k(T)$. Pair state $(k\uparrow,-k\downarrow)$
occupancies enter, with occupied and empty probabilities 
${\cal V}^2_k$ and ${\cal U}^2_k$ respectively given by
\bea
     {\cal V}_k^2 &=& \frac{1}{2}(1-\frac{\varepsilon_k}{E_k});~~
     {\cal U}_k^2  =  \frac{1}{2}(1+\frac{\varepsilon_k}{E_k});\nonumber \\
%  n_k&=&{\cal U}_k^2 f(E_k) + {\cal V}_k^2[1-f(E_k)] \nonumber \\
   n_k&=& {\cal V}_k^2 + \frac{\varepsilon_k}{E_k}f(E_k).
\eea
The single particle $k$ state
occupancy $n_k$ includes the pair occupancy ${\cal V}^2_k$ and
 QP excitations;
$f(E)$ is the Fermi-Dirac distribution of QPs.
In response to $\vec A$, their tensor result, provided here along axis $j$ 
for orthorhombic or higher symmetry, is\cite{Chandra1993} in analogy with
GL theory
\bea
\vec J^{BCS}_{s,j}(\vec r,T)&=&-\frac{e^2}{c} \sum_k 
                            \big[-\frac{\partial n_k}{\partial\varepsilon_k}
                        + \frac{\partial f(E_k)}{\partial E_k}\big]
                      \vec v_{k,j}^2 \vec A_j(\vec r) \nonumber \\
  &\rightarrow& -\frac{e^2}{c}N(0)v^2_{F,j} \nonumber \\ 
   &\times&  \Bigl[1-2\int_{\Delta}^{\infty}[-\frac{\partial f(E)}{\partial E}]
              \frac{E~dE}{(E^2-\Delta^2)^{1/2}}\Bigr]\vec A_j   \nonumber \\
 &\equiv& -\frac{c}{4\pi}\frac{1}{\lambda^2_{band}(T)}\vec A_j(\vec r).
\label{eqn:BCSsupercurrent}
\eea

The first thing to notice is that the T=0 value of both terms depend
solely on normal state properties. Next is that
this expression, with the arrowed simplification holding for an isotropic gap, 
gives (1st term) a \underline{diamagnetic} (London-like) 
contribution from the condensate; however, its magnitude is 
T-independent\cite{Chandra1993} 
and as mentioned  involves {\it only normal state} properties:
density of states $N(0)$, Fermi velocity $\vec v_k$.
There is also a \underline{paramagnetic} opposing QP current, 
monotonically increasing in magnitude with temperature\cite{Chandra1993} 
and sensitive to $\Delta_k(T)$. Both magnitudes scale as $N(0)v_{F,j}^2$.
This BCS `supercurrent' is, more correctly, the net current (super minus QP) in the SC state. 
The growing QP contribution with increasing $T$
opposes the supercurrent and equals it, thereby canceling it, at T$_c$.  
All the while, the 
supercurrent continues in shorting out electrical resistance.

A further feature of the BCS result is that the supercurrent
Eq.~\ref{eqn:BCSsupercurrent} must be solved 
together with the self-consistent gap equation,\cite{BCS1957} generalized to include band
indices being
\bea
\Delta_{kn}(T)=-\sum_{k'n'}V_{kn,k'n'}\frac{\Delta_{k'n'}(T)}{ 2E_{k'n'} }
   \tanh\frac{\beta E_{k'n'}}{2}
\label{eqn:gapeqn}
\eea
to obtain the $k,n$- and T-dependent gap, or the simpler energy dependent gap in the 
isotropic, single band limit. $V$ is the Fermi surface scattering matrix element.
Already with constant $\Delta(E)$ (no wavevector
or band dependence) and simplifications of the interaction matrix elements
the integral in Eq.~\ref{eqn:BCSsupercurrent} is a numerical task. 
Anisotropy and multiband effects in $\Delta(T)$ have gained
considerable attention especially in exotic superconductors, 
which lie beyond the scope of this discussion.
 
\subsection{Summary: progression of theory}
London theory established the linear relation between the supercurrent density and
the vector potential as being essential to explain the Meissner effect. 
It also introduced a seemingly crucial ``superfluid 
density'' $n^L$, and an unspecified mass $m^L$. The GL and BCS 
results supersede the London form respectively by (1) GL: its order parameter
gives a $\vec r$ and $T$ dependence to the $\vec J_s$ - $\vec A$ proportionality, more
involved and less intuitive than the London  penetration depth result, plus a new
term arising from phase softness and also proportional to $\vec A(\vec r)$ 
or boundary conditions, further complicating the 
analogy with the London penetration depth,   
then (2) BCS: QP contributions appear in addition to, and competing with,
the superfluid contribution
to the current density in the SC state, leading to further generalization
of the $\vec J_s$ - $\vec A$ proportionality. 
Chandrasekhar and Einzel\cite{Chandra1993} noted that charge conservation requires an 
additional, more involved `backflow' term in their BCS expression, which they 
provide, but becomes important only in highly anisotropic or multiband SCs. 
The $\vec r$ dependence in $\vec J_s^{GL}$ is not incorporated 
into the BCS result at this uniform SC state level of treatment. 

Comparison of
the GL and BCS bulk expressions\cite{Ketterson1999} leads to 
$\Delta(T)\propto |\Psi(T)|$, suggesting a generalization of BCS theory to 
$\Delta(\vec r,T)$ as implemented in the Usadel 
equations\cite{Usadel} for materials boundaries, or with reduced or no 
periodicity including strongly non-crystalline structure. 
GL theory provides no quantification of $|\Psi|^2$ in terms of its
phenomenological parameters.

\section{Drude weight}
\label{sec:Drude}
\subsection{Definition from the conductivity}
Using the Kubo linear response expression\cite{Kubo} and the Boltzmann
quasi-classical thermal distribution function $F(\vec r,\vec k;T)$ to account
for $\vec E$ and $\vec B$ fields and a thermal gradient $\nabla T$,
the expression for the current density becomes, for no $B$ field or
thermal gradient,
\bea
\vec j = -\frac{e}{V}\sum_k \vec v_k F(\vec k)=\sigma\vec E,
\eea 
defining the conductivity $\sigma$. 
A variational {\it ansatz} for an electric field alone takes the form
\bea
F(\vec k) &=& f( \vec k+\frac{e}{\hbar}\vec E \tau_k ) \nonumber \\
   &\rightarrow& f(\vec k)+e \vec E \cdot \vec v_k
     \left(\frac{\partial f}{\partial\varepsilon_k}\right)\tau_k
\eea
in terms of the Fermi-Dirac distribution $f(k)\rightarrow f(\varepsilon_k)$, 
the quasiparticle dispersion $\varepsilon_k$
and its gradient, the velocity $\vec v_k$. The effect of the $E$ field 
in steady state is to displace the Fermi surface, appearing in the Fermi-Dirac
distribution. By time-reversal symmetry $\varepsilon_{-k}=\varepsilon_k$,
$\vec v_{-k}=-\vec v_k$ so there is no current from the $f(\vec k)$ term. 

The expression for the conductivity becomes
\bea
\sigma_{xx}=\frac{e^2}{N}\sum_k v_{kx}^2 \tau_k 
    \left(-\frac{\partial f}{\partial\varepsilon_k}\right)
  \rightarrow e^2 N(0)v^2_{F,x}\tau
\label{eqn:conductivity}
\eea
with evident extension to the full tensor expression if desired. $N$ is the 
number of $k$-points in the sum.  The last expression replaces the
derivative of the Fermi function with a $\delta$-function as conventional, and 
replaces $\tau_k$ by its Fermi surface average $\tau$, 
referred to as the constant relaxation time approximation. Thereby a
simple expression is obtained in terms of band structure quantities and $\tau$.
This leads to the common use of the Drude plasma frequency $\Omega_p$
as
\bea
\sigma_{xx}(T)&=& \frac{\Omega_{p,xx}^2 \tau(T)}{4\pi}, \nonumber \\
  \Omega^2_{p,xx}&=& 4\pi e^2 N(0)v_{F,x}^2.
\label{eqn:Drude}
\eea 

The instructive manipulation of the textbook expression is as follows.
Integration by parts of the expression in Eq.~\ref{eqn:conductivity}
brings about the transformation\cite{PBA2006,Allen1988}
\bea
\frac{1}{N}\sum_k v_{k}^2 \left(-\frac{\partial f}{\partial \varepsilon_k}\right)
%%     \nonumber \\
 = \frac{1}{N}\sum_k \left(\frac{\partial^2\varepsilon_k}{\partial(\hbar k)^2}\right) f_k
   \equiv \frac{n_{v}}{m_{v}}.
\label{eqn:partialintegral}
\eea
The sum in the second expression is over {\it all occupied} (presumed
valence) states: the integral over $f_k$ would give $n_v$, and it is
weighted by the inverse mass. The
result is the valence electron density $n_{v}$ divided by the average
band mass $m_{v}$ over the entire valence bands. For this particle mass there
is massive cancellation of positive and negative masses, which for
each filled band vanishes.  Equation~\ref{eqn:conductivity} states that
the result is the Fermi
surface integral, which has led to observations in the literature
to make the connection
\bea
N(0)v_F^2 \leftrightarrow \left( \frac{n}{m} \right)_{eff}. 
\eea
In 1900 Drude\cite{Drude1900} had used $n/m$ in his
introduction to conductivity, as did the Londons (1935), 
as the Fermi statistics of electrons was
not well integrated into the theory at the time. 

This modernization may
have been first introduced in 1948 by F. London,\cite{London1948} where he pointed
out that the relevant quantities are $N(0)$ and $v_F$.
The ``Cooper pair'' term in $\lambda^{-2}_{band}(0)$,
Eq.~(\ref{eqn:BCSsupercurrent}), giving the
$T$=0 supercurrent-vector potential relationship, is the one of relevance here
\bea
%\frac{n_s^L(0)}{m_s^L}&\rightarrow& \left(\frac{n_s}{m_s}\right)_{eff}
%                       \rightarrow N(0)v^2_{F,j},\nonumber \\
   \frac{c^2} {\lambda^2_{band,j}(T=0)} &=& 4\pi e^2 N(0)v^2_{F,j}\equiv
       \Omega^2_{p,j}
\label{eqn:penetration}
\eea
in terms of the $j$ axis Drude plasma frequency\cite{Allen1971} $\Omega_{p,j}$.
No superfluid density, or any density, appears, only the availability of
carriers $N(0)$ and their velocity $v_F$. Interpreting Eq.~\ref{eqn:partialintegral},
the particle/unit-volume number density $n_s^L$ is supplanted by the
density of states in energy per unit-volume $N(0)$, while a second $
k$-derivative ($1/m^L$) is replaced by two first $k$-derivatives of energy ($v_k^2$).

\subsection{Penetration depth}
\label{sec:penetration}
A prominent method of measuring the magnetic penetration
depth has been from transverse field muon spin relaxation
TF-$\mu$SR, where the penetration depth profile is measured within,
and complicated by, the vortex lattice
produced by the magnetic field between $H_{c1}$ and $H_{c2}$. 
Brandt provided\cite{Brandt1988} a discussion and updated analysis
on the connection between $\lambda$ and the magnetic field
distribution sampled by isolated polarized muons in the flux
lattice of a Type II SC.\cite{Brandt1988} Considerations involve (i)
the Abrikosov index $\kappa$ reflecting vortex density,
(ii) interactions of vortices, (iii) imperfections of the vortex
lattice, primarily due to flux pinning or proximity to grain
boundaries, (iv) type (powder, crystal) and anisotropy of sample,
and (v) what assumptions of analysis are appropriate for a given
sample. An interesting comment from Brandt was that the ``true'' 
value of the penetration depth
$\lambda$ ({\it i.e.} for an isolated vortex) will be challenging to
pin down experimentally and may be a ``theoretical idealization.''\cite{Brandt1988}    
A description of muon physics broadly has been provided in the
`lecture notes' of Amato and Morenzoni.\cite{Amato2024}

As mentioned in the Introduction, most of the current interest 
than $\lambda$ {\it per se} is the
concept that has become to be known as a unitless ``superfluid density'' 
$\rho_s=[\lambda(0)/\lambda(T)]^2$. Its temperature
dependence should reflect the temperature dependence of the gap,
which in turn provides conditions on the order parameter.
Discussion, and results, remain presented in terms of 
$\lambda_L$ from the London picture.

Such studies are infrequent for low T$_c$, Fermi liquid SCs. One study 
by Hillier and Cywinski\cite{Hillier1997} from  $\mu$SR
spectroscopy is illustrative, with their analysis sometimes involving
homogeneous electron gas expressions. Their results for
two Fermi liquid systems (four materials) are relevant to the present
discussion. Using 25-30 atoms/nm$^3$ as a representative atomic density for a
multicomponent metal (more precise densities will not matter), their 
values of $n_s$ of several times $10^{29} m^{-3}$
convert to, for 
YB$_6$ (T$_c$=7.1K) and Y(Ni$_{1-x}$Co$_x$)$_2$B$_2$C 
($x$=0.00, 0.05, 0.10, with T$_c$=15K, 8.6K, 6.2K respectively), 
\bea
n_s&\sim& 1.0/atom:~\text{YB$_6$},\nonumber \\
n_s&\sim& 1.0/atom:~\text{Y(Ni,Co)B$_2$C},\nonumber\\
n_s&\sim& 0.5/atom,~\text{Y(Ni$_{0.95}$Co$_{0.05})_2$B$_2$C},\nonumber\\
n_s&\sim& 0.2/atom,~\text{Y(Ni$_{0.90}$Co$_{0.10})_2$B$_2$C}.
\eea
These are
{\it truly macroscopic values} of superfluid density, as it seems was
pictured by many at the time. These carrier densities being
involved in screening
would imply a highly {\it non-degenerate} electronic state, even at low T$_c$.
Other indications are that these materials are degenerate
Fermi liquid SCs, with that impact discussed in following sections.
A central interpretation was that $\lambda^{-2}$ is proportional to
$n_s$.  Some examples of penetration depths $\lambda_{band}$ are
provided in Table I, and discussed in Sec. III.B.2.

\subsection{Optical spectroscopy}
\label{sec:farIR}
The previous discussion has involved only steady state, ground
state properties: static screening and superfluid response to
a magnetic field. Optical response can be divided into far-IR
(intraband), interband transitions, or assigned to the 
$\delta$-function $\delta(\omega)$ response of the SC.
 Optical spectroscopy provides insights
into the topics of this paper.

\subsubsection{Conductivity sum rule, superfluid response}
Upon opening of the SC gap the electron spectral density below the gap energy
decreases (ideally to zero, giving the IR ``window'' of a BCS SC),
with the spectrum above being altered up 
to a few $2\Delta(T)$. As a result of the gap opening there is a loss
of spectral density -- missing area under the real part of the conductivity
$\sigma_1(\omega;T)$ of the conductivity. It was discovered soon after BCS
theory that the missing response condenses into a $\delta$-function
contribution $\propto\delta(\omega)$ that is connected with the 
infinite d.c. conductivity, {\it i.e.} the supercurrent.
This condensate response and optical properties more generally  have
attracted a great deal of interest related to superfluid properties.

The foundation of the spectral density approach is the 
electrical conductivity sum rule ($f$-sum rule)
which for any phase at any temperature is, see Kubo,\cite{Kubo}
\bea
\frac{2}{\pi}\int_0^{\infty} \sigma_1(\omega) d\omega=e^2\frac{n}{m}
\eea
in terms of the electron {\it total number} density $n$ (including core electrons)
and free electron mass $m$. Application 
of this sum is not useful in this precise form due to the impractically, if
not impossibility, of extending optical studies to energies well above
that where the most strongly bound (core) electrons are excited into the continuum. 

The usual procedure is to restrict the integral to an upper limit
$\omega_m$ giving $n(\omega_m$) (or $n(\omega_m)/m(\omega_m))$ on the 
right hand side, finding an optimum frequency of separation of intraband with 
transitions including interband.  The
practice is to extend the integral to variable $\omega$
and to make an optimal choice.
The mass $m_{\omega_m}$ would then be an average over the Fermi surface,
with $n_s(\omega_m)$ providing the carrier density participation in
intraband transitions: 
\bea
\int_0^{\omega_m} \sigma_1(\omega) d\omega
     =\frac{1}{8}4\pi e^2\frac{n(\omega_m)}{m(\omega_m)}
\label{eqn:partialsumrule}
\eea  
for the chosen cutoff.
The notational similarity to the London $n/m$ ratio is much of the discussion,
while the fraction 1/8 has a formal origin. This expression
is suggestive of an effective Drude plasma frequency $\Omega_{p,eff}(\omega_m)$.

\subsubsection{{\rm far}-IR spectroscopy}
This subsection will focus on the low frequency
(Drude) regime, together with the $\delta$-function response discussed
above, with emphasis given to
the difference between the normal and SC states.
Basov and Timusk\cite{Basov2005} reviewed the two-fluid aspect
(normal, superfluid) of the optical conductivity in the far-IR regime,
expressed in terms of the expression  
\bea
\sigma_1(\omega,T)&=&\frac{1}{8}\frac{4\pi e^2 n_L}{2m_L} \Big[w_n(T)
    \frac{\tau}{1+\omega^2 \tau^2} + w_s(T)\delta(\omega)\Big] \nonumber \\
  &\rightarrow& \frac{(\Omega_p^*)^2}{8}~\Big[w_n(T)
    \frac{\tau}{1+\omega^2 \tau^2} + w_s(T)\delta(\omega)\Big],
\eea
with fractions satisfying $w_n(T)$+$w_s(T)$=1 for all $T$. 
The second expression converts the
prefactor into the Fermi surface  property 
$\Omega_p$ via Eq.~(\ref{eqn:Drude}), reflecting that lowest energy excitations 
are confined to scattering
along the Fermi surface. The presumption is that the scattering
around the Fermi surface of the excited quasiparticles (with
scattering time $\tau$) is independent from the
response of the condensate, {\it i.e.} the normal state $\tau(T)$ is applicable,
and that it is frequency independent. Here one can note that the BCS
expression Eq.~\ref{eqn:BCSsupercurrent} suggests that, in screening, the QP current is
short-circuited by the supercurrent conductivity, leaving no 
scattering time $\tau$ for the QPs in the SC state.  

The first term is due to the (broken Cooper pair) quasiparticles,
with corresponding fraction $w_n(T)$.  The second is the fraction and weight of the
$\delta$-function response of the condensate, with magnitude $(\Omega_p^*)^2/8$
at $T$=0.  $\Omega_p^*$ is the experimental Drude plasma frequency, including dynamical
corrections.

At the next level of theory, account would need
to be taken into the spectral difference between normal and SC states.
Thus 
\bea
 \int_{0+}^{\bar{\omega}}
       \bigl [\sigma_1^n(\omega,T\rightarrow 0)
             -\sigma_1^s(\omega,T\rightarrow 0)\bigr]d\omega
  \approx \frac{1}{8}(\Omega_p^*)^2
\eea
is the weight in the $\delta$-function response. $\bar{\omega}$ is a somewhat ambiguous
limit, the idea being that superconductivity and normal  
conductivities have converged  
before interband transitions become appreciable.
The experimental constant has conventionally been denoted $\omega_p$ (not to be
confused with the high frequency
optical plasma frequency). Thus the experimental value should be identified as 
$\omega_p\leftrightarrow \Omega_p^*$. 

%\subsubsection{Interactions}

{\it Effect of interactions.}
Silin specified a parameter\cite{Silin1,Silin2} $\cal{A}$ (using the
notation of Ref.~[\onlinecite{WEP1975},\onlinecite{WEP1976}]) that quantifies
the difference between theory and experiment due to interaction effects
 left out of the theory:
\bea
 (\Omega_p^*)^2 = \Omega_p^2(1+{\cal A}).
\eea
This interaction strength
parameter quantifies the change in penetration depth due to interactions:
(1) electron-electron interactions from
viz. Landau's theory of Fermi liquids, or from more modern methods,
(2) electron-phonon (el-ph) mass enhancement, and
(3) other interactions such as spin fluctuations. 
A few values of ${\cal A}$ are given below where comparison with experiment
is available.
%From the above
%data, ${\cal A}$$\approx$0.68 (0.80) for Nb (Mo), with relatively large
%uncertainty for Nb.

Allen described\cite{Allen1971} how electron-phonon coupling
(intraband scattering) affects
the far-IR optical constants (Coulomb effects are expected to be negligible
for the mass enhancement in such low T$_c$ metals).
The primary effect is the  electron-phonon mass enhancement by $1+\lambda_{ep}$,
increasing $N(0)$ by this factor but decreasing $v_F$ by the same
factor. Since $v_F$ is squared,
\bea
(\Omega^*_{p})^2 = \frac {\Omega^2_{p}} {1+\lambda_{ep}}.
\eea
Consequently, the penetration depth is increased by $\sqrt{1+\lambda_{ep}}$
by interactions. Heuristically,
screening of the field is decreased by this factor. Equating the two expressions
for $(\Omega_p^2)^*$ gives, for these Fermi liquid metals,
\bea
{\cal A} = -\frac{\lambda_{ep}} {1+\lambda_{ep} },
\eea
so ${\cal A}$ is negative and in the neighborhood of -0.35 to -0.70 for known
superconductors. A few numbers will be given below. 

\subsection{Examples: theory and experiment} 
{\it Group VB-VIB transition metals.}
For elemental 
Nb (T$_c$=9K), 
Mo (T$_c$=0.9K), 
Ta (T$_c$=4.5K), and
W (T$_c$=0.015K),  
values reported by Chakraborty {\it et al.}\cite{Bulbul1976}
of $N(0)$, $v_f$, and $\hbar\Omega_p$ from Eq.~(\ref{eqn:Drude}) are 
reproduced in Table I.
The band values for Nb (Mo) are $\hbar\Omega_p$=8.9 (8.2) eV,
reflecting how the greater than a factor of two values of $N(0)$
are strongly compensated by the squared-velocity factor.
These SCs have two Fermi surfaces, and many multicomponent
SCs have more. A recent study for Nb finds\cite{Zarea2023} that
its two Fermi surfaces have average gaps that differ by 20\%, 
each with an anisotropy around 2\%. This gap difference is not
uncommon in intermetallic superconductors, it is treated routinely (if approximately)
by Eliashberg theory codes, and such small distinctions are difficult
to resolve in experiment.  
One relevant factor is that these Fermi surfaces are a large majority
$4d$ character, thus very similar in properties (not $s$ and $d$ as
in early suggestions), and the cubic
structure also reduces anisotropic effects. 
The corresponding penetrations depths are nearly identical for Nb and Mo
($\sim$33 nm) and somewhat larger and again equal for Ta and W ($\sim$50 nm).
The strong cancellation of $N(0)$ with $v_F$ 
suggests that band values will not differ greatly among relatively low 
T$_c$ Fermi liquid transition metal compound SCs.

%\begin{widetext}
%\begin{singlecolumn}
%\begin{center}
%\begin{table}[!h]
\begin{table*}[ht]
\caption{Values of the various properties discussed in the text,
for five BCS singlet superconductors from weak to strong coupling,
low T$_c$ to high T$_c$. 
Band structure numbers are averages over several Fermi surfaces.
High pressure values are: SH$_3$, 200 GPa;\cite{Ghosh2019} 
LaH$_{10}$, 250 GPa.\cite{Heil}
Values for orthorhombic LaNiGa$_2$\cite{Botana} and
LaNiC$_2$\cite{Subedi2009} are for the 
$x,y,z$ axes respectively. For comparison, the r.m.s axis 
dependent ($j=x,y,z$)
values are quoted for the cubic compounds: $v_{F,j}=v_F/\sqrt{3}$.
2$\Delta_o$=3.5-4 $k_BT_c$ gives the listed values. For SH$_3$ the
derived (BCS) gap agrees well with the experimental value.\cite{Du2025}}
%\begin{center}
\begin{tabular*}{\textwidth}{l|c|c|ccc|c|c||rrr||}
 \multicolumn{1}{c|}{Material}&
 \multicolumn{1}{c}{T$_c$}&
 \multicolumn{1}{c}{$v_{F,j}$}&
 \multicolumn{1}{c}{$N(0)$}&
 \multicolumn{1}{c}{$N(0)$}&
 \multicolumn{1}{c}{$\bar{N(0)}$}&
 \multicolumn{1}{c}{$\hbar\Omega_{p,j}$}&
 \multicolumn{1}{c}{$\lambda_{band}$}&
 \multicolumn{1}{c}{$2\Delta_o$}&
 \multicolumn{1}{c}{${\cal N}_s=N(0)\Delta_o$}&
 \multicolumn{1}{c}{${\cal N}_{atom}$}
   \\
\multicolumn{1}{c|}{}&
 \multicolumn{1}{c}{(K)}&
 \multicolumn{1}{c}{$(10^{7}$cm/s)}   &
 \multicolumn{1}{c}{(eV f.u.)$^{-1}$} &
 \multicolumn{1}{c}{(eV$^{-1}$   nm$^{-3}$)} &
 \multicolumn{1}{c}{(eV$^{-1}$ atom$^{-1}$)} &
 \multicolumn{1}{c}{(eV)}&
 \multicolumn{1}{c}{(nm)}&
 \multicolumn{1}{c}{(meV)}&
 \multicolumn{1}{c}{($10^{-3}$nm$^{-3}$)}&
 \multicolumn{1}{c}{(el./atom)}
   \\ \hline
%%%%          Tc    v_F,j    N(0) N(0)  N(0)      hOmega  lambda  2\delta  N_s      N_s
Nb        &~9.2  &~3.5      &~1.40&~81&1.40&      6.2  &    32  &2.65&    107& 1.8$\times 10^{-3}$\\
Ta        &~4.5  &~5.0      &~1.32&~76&1.32&      9.4  &    49  &1.30&     49& 0.8$\times 10^{-3}$\\
\hline
Mo        &~0.9  &~3.9      &~0.62&~34&0.62&      5.9  &    34  &0.26&   4.4& 0.8$\times 10^{-4}$\\
W         &~0.015&~5.5      &~0.55&~30&0.55&      8.8  &    51  &0.04&   0.6& 0.1$\times 10^{-4}$\\
\hline
\hline
LaNiGa$_2$&~2.0 &2.0,1.5,2.3&~3.66&~23&0.91&2.7,2.0,3.1&52,68,45&0.60&    12& 2.3$\times 10^{-4}$\\
LaNiC$_2$ &~2.7 &1.8,1.5,1.5&~2.60&~23&0.65&2.4,2.0,2.0&46,68,29&0.81&    ~9& 1.7$\times 10^{-4}$\\
\hline
\hline
SH$_3$    &~200 &~1.4       &~0.63&~48&0.18&      1.9  &    99  &  62&  1550& 5.6$\times 10^{-3}$\\
LaH$_{10}$&~250 &~0.87      &~0.88&~31&0.088&     0.95 &   206  &  85&  1320& 3.5$\times 10^{-3}$\\
  \hline
\end{tabular*}
\label{table1}
\end{table*}
%\end{center}
%\end{singlecolumn}
%\end{widetext}

The high T$_c$ for an elemental metal has focused more attention on Nb.
From three far-IR (at the time) studies the mean value\cite{explainNb} 
was $\hbar \Omega^*_{p}$ around 6.1 eV. 
The experimental value\cite{3references} for Mo is 5.9 eV.
%Theory-experiment comparison give values of the renormalization of
%the ``optical mass'' (jargon of the time) of ${\cal A}= -0.5\pm 0.2$
%(-0.5) for Nb (Mo).
With $\lambda_{ep}$=1.05 [\onlinecite{Zarea2023}] ($\sim$0.5) for
Nb (Mo), the el-ph contribution to the Silin parameter\cite{Silin1,Silin2} ${\cal A}$ would be
=- 0.50 (-0.33) for Nb (Mo). These values indicate that electron-phonon coupling
accounts for most of the renormalization of $\Omega_p$ to $\Omega_p^*$.

Experimental values of penetration depth provide further comparison.
Using resonant diode frequency shifts due
to field penetration into the SC sample, Maxfield and McLean
reported\cite{Nb-lambda1965} for Nb a penetration depth of
$\lambda_{exp}\approx$ 47 nm. This value is comparatively
close to the band value of Table I, 32 nm, especially considering that
interactions will increase the penetration depth over the band value. More recently,
McFadden {\it et al.} have performed a more direct measurement\cite{McFad2025}
of the magnetic screening profile $B(z)$ in Nb
using low energy muon spin spectroscopy, thereby
sampling the magnetic field at various depths. Their resulting penetration
depth was 29 nm, indistinguishable from the band value.
A similar earlier study by Suter {\it et al.}\cite{Suter2006} found
$\lambda$$\sim$50 nm for Ta, again indistinguishable from the band value in Table I.
These results 
suggest that the electronic structure results should serve as a
useful  anchor in providing the magnitude of $\lambda$.
The following example complicates this picture.

{\it LaNiGa$_2$}, based on most properties,\cite{LaNiGa2a,LaNiGa2b,Ghimire2024} 
provides a recent example of an orthorhombic
intermetallic low T$_c$ SC ($T_c$=2K), Fermi-liquid-like and singlet paired 
based on normal and SC properties, but it has been proposed 
as a triplet SC based on zero-field $\mu$SR depolarization data.\cite{LNG1,LNG2}
Unlike the Type I superconductors Mo, Ta, W, 
and elemental Nb on the borderline of Type I / Type II, LaNiGa$_2$ is clearly
Type II (although different axes lead to different values of the
Abrikosov parameter $\kappa$, all lie within the Type II regime).
The directional band velocities $v_{F,j}$\cite{Botana}
are $1.9\pm 0.4$$\times 10^7cm/s$, clear but modest anisotropy in
a 3D metal. The band penetration depths are 53$\pm$11 nm, roughly a
factor of two larger than Nb and Mo, but similar to Ta and W. See Table I for
the axis-dependent values.
%The value of the
%condensate density ${\cal N}_s$ is, in per atom values, in the general
%range of Mo, which has a similar low T$_c$.

Single crystals of LaNiGa$_2$ led to axis-dependent values 
of Badger {\it et al.}\cite{LaNiGa2a}
of the  GL penetration depth\cite{LaNiGa2a,LaNiGa2b} obtained
from the conventional application of the GL theory expressions
for $\lambda_{GL}$ and the coherence length $\xi$
(using zero temperature, axis specific quantities)
\bea 
\lambda_{j,GL}&=&\Phi_o/2\sqrt{2}\pi H_{c}(0)\xi_{j,GL},\nonumber \\
      H_{j,c2}&=&\Phi_o/2\pi\xi_{j,GL}^2,   
\eea
with critical fields available from experiment. 
Note that the GL expressions giving $\lambda_{GL}$ seem to have little in
common with the electronic structure expression Eq.~\ref{eqn:Drude}.
The GL ($a,b,c$) values $\lambda_{GL}=(174,509,185)$nm are
3.5-7 times larger than $\lambda_{band}$, see Table I.
From $\mu$SR data Hillier {\it et al.}\cite{LNG1}
reported 350(10) nm for polycrystalline
samples, a reasonable average of the Badger {\it et al.} values.
The discrepancy between band and experimental values remains.
Some of this discrepancy may be due to the Silin dynamic interaction
effects.

This band versus GL difference
is apparently the scale of the difference between GL theory
and the electronic structure prediction for intermetallic superconductors. 
In this vein, another comparison is that between the BCS expression for
$\xi_j=\hbar v_{F,j}/{\pi\Delta_o}$=(24,18,28)nm whereas the GL
values\cite{LaNiGa2a} are (52,18,47)nm, differing by no more than a factor of two,
hence within all of the uncertainties. 
It has been reminded\cite{Ghimire2024} that the BCS value for the coherence length 
generally differs from the GL value. The difference has some importance because
the band values suggest LaNiGa$_2$ is somewhat nearer the Type I/Type II
boundary whereas the GL values lie further into the Type II regime. 

A comparable compound would be LaNiC$_2$, with a related crystal structure
to LaNiGa$_2$ but without inversion symmetry, and with similar T$_c$=2.7K. 
As for LaNiGa$_2$, LaNiC$_2$ was proposed to show time-reversal symmetry
breaking at T$_c$, based on $\mu$SR depolarization measurements.\cite{Hillier2009} Values
from $H_{c2}$ for polycrystalline samples suggested $\lambda$$\sim$125 nm,
with a value for the $a$-axis being 155 nm. These values
are roughly $\sim$2 times smaller than the
GL values of LaNiGa$_2$. A DFT study of LaNiC$_2$ pictured the variation
of $v_F$ over the Fermi surface\cite{Subedi2009} and provided the axis-projected velocities,
with smaller variation than in LaNiGa$_2$ (Table I) and with much
smaller values of $\lambda$.

{\it Compressed metal hydrides} provide
an extreme example of Fermi liquid SC, with SH$_3$ and LaH$_{10}$
displaying  T$_c$= 200K and 250-260K respectively,
for pressures in the 160-200 GPa range. They have been proposed
to be considered
as hydrogen superconductors,\cite{Quan2019} the metal atom being
effectively a bystander, at least for determining T$_c$.  The occupied
 H valence bands are 25 eV wide (total bandwidth even larger)
due to the compression. This bandwidth suggests a new
class of Fermi liquid SCs, different from ambient SCs, but still described
by Eliashberg theory.\cite{Pickett2023} The Drude plasma energy, from Fermi
surface values for SH$_3$ from Ghosh {\it et al.},\cite{Ghosh2019}
and from Ferreira and Heil\cite{Heil} for LaH$_{10}$ respectively, are
$\hbar\Omega_p=0.95$ eV (1.9 eV). These values lie at the
low end for SCs at ambient pressure, 
but roughly a conventional value of $\Omega_p$.
Generally, it appears that electronic structure values of
$\hbar\Omega_p$ for Fermi liquid metals will not differ much from
the 2-9 eV range, so penetration depth values of $\lambda_{band}$ even
for multiband metals at ambient pressure 
are likely to cluster within less than one order magnitude.
\vskip 2mm \noindent

\subsection{Condensate density}
\label{sec:condensate}
\subsubsection{Definition of ${\cal N}_s$}
The BCS result above separates the penetration depth $\lambda_{band}(T)$
-- the proportionality between $\vec J_s$ and $\vec A$ -- from any particle
density. As mentioned, ``superfluid density'' has intrigued the field
for decades, so the question arises: is there a definition of some
related scalar density, and
does it have any place in determining SC properties?

Phenomenologically, one might expect a {\it condensate density} ${\cal N}_s$
at zero temperature to be the number density of normal state
electron states (of both spins) absorbed into
the condensate
\bea
{\cal N}_{s}(T=0)=N(0)\Delta_o,
\eea
{\it i.e.} only the states
from $-\Delta_o$ to 0 were occupied, hence the energy width here is
$\Delta_o$ rather than the full gap $2\Delta_o$ ($N(0)$ is for both spins).
Combining equations (2.45) and (3.29) of the
BCS paper\cite{BCS1957} gives their value from their density of Cooper pairs,
\bea
{\cal N}_s^{BCS}(T=0)\approx 2k_BT_c N(0) \approx N(0)\Delta_o
\label{eqn:condensate}
\eea
for the density of paired electrons (twice their density of Cooper pairs)
in the BCS wavefunction, the same result.
Here it has been incorporated that $2\Delta_o\approx$ $3.5$-$4$~$k_BT_c$ for
weakly to moderately coupled superconductors as covered by the BCS
wavefunction.
Extension of the BCS expression for the temperature dependence of
the gap suggests a
T-dependence ${\cal N}^{BCS}_s(T)=N(0)\Delta(T)$
that should be instructive for those of interest in its temperature
evolution.

For comparison, thermally excited electrons available for
low energy normal state transport or screening are those
in the energy range $\sim\pm k_BT$ relative to $E_F$, which at T$_c$ is
\bea
{\cal N}_n(T_c)\approx 2k_BT_c N(0)\approx {\cal N}_s^{BCS}(0),
\eea
{\it i.e.} all thermally excited electrons in the normal state at $T_c$
finally enter the condensate at $T$=0. This correspondence
between normal and SC carriers is not widely recognized,
and may deserve further scrutiny.

Some specific observations are instructive.
Values of $N(0)$, $\Delta_{\circ}$, and ${\cal N}_s$  are among
the data displayed in Table I for
the representative Fermi liquid SCs  
Nb ($\lambda_{ep}$=1.05) and Mo ($\lambda_{ep}$$\sim$0.5), 
lie at the extremes of elemental transition metal SCs,
Nb having the highest T$_c$, Mo an order of magnitude
lower. The larger $N(0)$ by a factor of 2.4 in Nb leads to higher T$_c$
hence $\Delta_{\circ}$ by a factor of 10, and results in a factor of 
27 difference in their condensate densities
${\cal N}_{s}$. The condensate density is thus a superlinear function
of $N(0)$. For a physical picture, there is one SC electron in each
direction for 6 atomic distances for Nb, whereas it is for Mo 
a radius of 20 atomic distances in Mo. 

Orthorhombic LaNiGa$_2$ 
has a 30\% larger value of $N(0)$ compared to Mo (on a per unit volume basis) 
but more than twice as large T$_c$ and $\Delta_{\circ}$, leading to a factor of 2.7 larger
condensate density. At the other extreme are the first two metal 
hydrides discovered to have very high T$_c$
at very high pressure, SH$_3$ and LaH$_{10}$, with $N(0)$ 25\%
larger for SH$_3$. The values of $\Delta_{\circ}$ ({\it i.e.} T$_c$) are close but
reversed (T$_c$ is not so directly connected with $N(0)$ in these hydrides), 
leading to rather similar values of absolute ${\cal N}_s$ for these two,
but much larger than for ambient temperature SCs.

\subsubsection{Giving meaning to ${\cal N}_s$}
While the Gorter-Casimir separation\cite{Gorter1934} of free energies into those from
superconducting and normal densities are challenged by 
BCS theory,\cite{Schrieffer1964}
one can address the change in the BCS grand canonical potential 
${\cal G}(T)=E(T)-\mu{\bar N}$ with energy $E(T)$, via ${\cal N}_s$. 
From temperature $T$ to $T+dT$ below T$_c$, the increment 
is 
\bea
\delta{\cal G}=2\Delta(T)\frac{d{\cal N}_s(T)}{dT}dT=
 2N(0) \frac{1}{2}\frac{d\Delta^2}{dT} dT,
\eea
because each new Cooper pair at temperature $T$
gains an energy $2\Delta(T)$.
Integrating from $T_c$ to $T$, and
introducing the condensate (Cooper pair) density 
${\cal N}_s^{pair}$=${\cal N}_s/2$ of BCS, the gain in SC energy is
\bea
 \Delta {\cal G}(T)
%    &=&\int_T^{T_c}2\Delta(T) \frac{d{\cal N}_s^{pair}(T)}{dT}dT\nonumber  \\
    &=& 2N(0)\int_T^{T_c} \frac{1}{2}\frac{d\Delta^2(T)}{dT}~dT\nonumber \\
    &=&-2N(0)\Delta^2(T) =-{\cal N}_s^{pair}(T)\Delta(T).
\eea
The BCS estimate\cite{BCSestimate} of the gain in energy due to additional pairing
at temperature $T$ is, for 
an isotropic gap $\Delta(T)$,
\bea
\delta[E-\mu \bar{N}]&=&-\frac{1}{2}N(0)\Delta^2(T)=-\frac{1}{2}{\cal N}_s(T)\Delta(T),
    \nonumber \\  
                    &=&-{\cal N}_s^{pair}(T)\Delta(T),
\eea
the same result, and assigning a net gain of {\it mean energy} $\Delta(T)$ 
from each member of the Cooper pair added from $T_c$ to $T$. Note that 
the above expression for $\delta{\cal G}(T)$ is heuristic, whereas
the BCS result is a calculated estimate of this non-perturbative
energy difference that would include pair-pair interactions.  
 
Clarification of the role of the condensate density in superconducting 
properties may help in furthering both understanding and applications. Most
of the use of this concept has focused on a superfluid density $n_s$ 
in relation
to applications limited by critical supercurrent densities, while not
being able to quantify this superfluid density magnitude.\cite{Beasley2011}

Cuprate superconductors are widely discussed as limited due to
``low Cooper pair density.'' The condensate density defined here 
 for Fermi liquid SCs depends on available states $N(0)$, whose band value
 is relatively low, times
the half-gap $\Delta_o$, which is large. The critical supercurrent density
in an electric field should still involve an interacting value 
$N^*(0)$ and the carrier velocity, combined with complications that include the 
self-field, defects, vortex dynamics, etc. of the superconductor. The
precise way in which these properties depend (or not) on the condensate density
${\cal N}_s$ could help to clarify the limitations and possibilities.  

\section{Exotic Superconductors}
\label{sec:exotic}
The previous discussion has been aimed at Fermi liquid metals,
that is, metals that are degenerate in the sense that all states
below $\sim(E_F-k_BT)$ are Fermi-blocked from participating in 
transport or far-IR dynamics at temperature $T$ or below,
and that excitations are Fermi liquid quasiparticles. Even heavy 
fermion metals satisfy this criterion, although the temperature 
below which this applies is closer to the kelvin level rather than 
the $10^3$K level.

The high T$_c$ cuprates, which are doped 
in relation to some reference valence electron level, provide the 
notorious metallic counterexample to this class of conventional
metals. Even so, optimally doped cuprates have been found to display well-defined 
Fermi surfaces essentially in agreement with standard electronic
structure theory.\cite{WEP1992} These Fermi surfaces are much applied in
modeling order parameter character. This observation implies filled and thus
Fermi-blocked states below $E_F$, with support from angle-resolved
photoemission spectroscopy. 

However, the excitations, in transport
and several other probes, do
not follow those of a Fermi liquid. It is this class that has
attracted the greatest attention to ``superfluid density'' so
a mention is warranted here.  This question has been studied 
experimentally by a few techniques, including 
muon spin resonance, rotation, and relaxation ($\mu$SR),
far-IR spectroscopy, a tunnel diode oscillator 
technique, and perhaps others. In all cases the underlying picture has been
in London terms, where the inverse square penetration
depth is proportional to $n_s/m_s$.
Three examples of organization of experimental studies will be 
mentioned, the references should be consulted for the manner of
analysis and more results that are not mentioned here.
For a review of the penetration depth in unconventional
SCs, see Prozorov and Giannetta.\cite{Prozorov2006}  This paper provides
an extended discussion of the tunnel diode oscillator experiment,
and discusses the various challenges to determining the absolute
penetration depth. 

{\it The Uemura plot}. 
An influential  study from transverse field $\mu$SR spectroscopy was presented by
Uemura {\it et al.}\cite{Uemura1989} for four classes of 1, 2, 3
layer cuprates at somewhat underdoping to optimal doping, finding
that around optimal doping T$_c$ was proportional to the muon depolarization rate,
expected to be proportional to $\lambda^{-2}$, hence from the London
picture proportional to the SC carrier density $n_s$. The result was
a linear relation between $n_s$ and T$_c$.
Using an estimated $m\sim 5m_e$ along with other analyses for these
cuprates, several values for optimally doped samples clustered around 
$n_s=2\times 10^{21}cm^{-3}$. This value is of the order of
0.1-0.2/plaquette, and as they noted, it is roughly
consistent with valence counting estimates of the hole doping, {\it i.e.} each
hole contributes entirely to the dynamics, without any
Fermi blocking of occupied states.

{\it The Homes plot.} 
Homes and collaborators\cite{Homes2004,Tu2010,Homes2022} analyzed
optical conductivity on several cuprates to provide a different 
but related update of the Uemura plot. The technique was to measure 
the reflectivity from a large region, viz. 10 meV to 5 eV, and use 
Kramers-Kronig relations to obtain the complex dielectric function
$\epsilon(\omega)$ in the low frequency Drude regime for analysis.
Extracting the Drude plasma frequency $\omega_p$ ($\Omega_p^*$
from previous sections of this paper) and applying the London
picture $\omega_p^2\propto n_s/m_s$ along with observed empirical
trends of the conductivity $\sigma(T_c)$, they obtained the relation
\bea
n_s\propto \sigma_{dc}(T_c)T_c
\eea
for a number of cuprate materials for $a$-$b$ plane polarization. 
The observation that in several cuprates $\sigma_{dc}(T_c)$ is
proportional to $T_c$ was incorporated into the analysis.

Several near optimally doped cuprates clustered around a density
that converts to the neighborhood of $10^{-5}$/plaquette. This value is four
orders of magnitude below the conventional doping density.
The difference may be in (1) different probes, (2) little actual
use of the density in the analysis, related to (3) an incomplete
microscopic theory to guide the analysis. Homes, like Uemura, had in mind the
London relation $\lambda^{-2}\propto n_s/m_s$, although the
absolute value of $\lambda$ was not given much attention. 

{\it The Tanner Rule.}
A different approach to connecting carrier densities to T$_c$ was taken by
Tanner and collaborators.\cite{Tanner1998} Like Homes {\it et al.}, they
used reflectivity data to extract complex $\epsilon(\omega)$, in a 
region from minimum energies of 4-20 meV to maximum of 0.5-4 eV,
depending on sample type and size, spectrometer, and temperature range.
A fundamental difference was that they found that Drude 
behavior was not observed in the expected (measured) regime.
Their analysis identified {\it three distinct carrier densities} from eight
samples from four cuprate classes, all near optimal doping.
The densities, all per Cu atom, were (i) an effective density $n_{eff}$
obtained from doping induced carriers (the commonly used measure),
(ii) a SC density $n_s$ as discussed here, and (3) a density of non-superconducting
Drude carriers $n_D$. These distinctions redirected
the interpretation of carrier densities.

The original paper should be consulted for specifics of the analysis,
which relied heavily on the partial sum rule Eq.~(\ref{eqn:partialsumrule}). The doping
count $n_{eff}$ ranged from 0.15-0.59/Cu. The primary conclusions were that 
(i) $m\sim 5m_e$ provided the most reasonable analysis, and (ii) the
fraction of doped carriers that superconduct is $n_s/n_{eff}$ =
20\%, with very little scatter. The separate densities show correlation
with T$_c$, while the ``Tanner rule'' of 20\% participation of doped
carriers into the SC state is constant over the eight samples studied.
The reported values of $n_s$ are of the same
order of magnitude as Uemura (the doping level), and unlike the small Homes values.  

Drawing correlations between T$_c$ and other measured properties remains a
continuing area of interest. For example, Taylor and Maple\cite{Taylor2009} incorporated the GL
condensation energy $E_s^{GL}\propto H_c^2$, interpreted by them as that
energy within the coherence volume of a Cooper pair, as a fundamental energy
scale. The ratio $n_s/n_n$, without specific definition
of either, with $\xi_{\circ}^{GL}$ and $m_s$ were components in their
relation for T$_c$. Rearranged, their expression gave a linear relation
between $H_{c2}$ and T$_c$. Their relation was presented for nearly 20
correlated electron superconductors (with MgB$_2$ as an example of
a multiband  SC), with
a roughly linear correlation, but sometimes with large uncertainty bars. Again,
the London picture of $n_s$ and $m_s$ played a role in their analysis.    

\begin{comment}
{\it Synopsis of this section.}
This 20\% fraction of superconducting carriers, together with
the better justified mass enhancement
of five, has become referred to as the ``Tanner rule.'' Given that optimally
doped cuprates have well defined (and much used in research) Fermi surfaces,
it might be suspected that the invisible 80\% of carriers not participating
in the superconducting state would be 
Fermi-blocked within the Fermi surface, thus non-dynamic. The excitations
in cuprates are of course not of Fermi liquid type. A definition of
carrier density with some rigor is the partial sum rule giving $n_{eff}(\omega_c)$
at the chosen cutoff. It is of interest that many Tanner values (usually
for self-doped cuprates) of
density $n_{eff}(\omega_c)$ in per-Cu units is much larger (0.40-0.55/Cu) than the
standard plot of the T$_c$ dome displays. [The Tanner values of
$n_s$ for the bilayer cuprates
clusters around 0.08-0.12 per Cu, while it is only 0.028 for the
optimally doped  (La,Sr)CuO$_4$ sample.]  
\end{comment}

\section{Summary}
\label{sec:summary}
As stated in the Introduction, the purpose here is to promote
a microscopic theory of the penetration depth that will stimulate
a more precise understanding of ``superfluid density'' and related properties.
It has been clarified (not at all a new result), 
for the Fermi liquid metals of focus, that the superconducting penetration depth 
$\lambda$, involving only static screening, is independent of any particle density,
depending only on particle availability $N(0)$ and Fermi velocity $v_F$.
It is through these properties, as embodied in the Drude plasma frequency
$\Omega_p$, that provides the scale of the penetration depth and thereby
the supercurrent density.
Clarifying some of the common language may affect the understanding 
of critical currents in superconducting applications.\cite{Beasley2011} 

There is a quantity,
the condensate density, twice the density of Cooper pairs, that is 
well defined from the BCS wavefunction and
for Fermi liquid superconductors at ambient pressure is on the order of
$10^{-4}$-$10^{-3}$  per atom.  It remains to be decided how the
magnitude of this, or the penetration depth, is altered by strong 
interactions in high temperature (cuprate, Fe-based) superconductors.
The superfluid $\delta$-function response, well known as as 
$\omega_p^2/8\rightarrow(\Omega_p^*)^2/8$, is determined by the same
Fermi surface quantities as the penetration depth.
The condensate density along with other Fermi surface parameters 
are provided for a few conventional superconductors, as well  as
for two highly compressed metal hydride superconductors.
Finally, this condensate density is the same density as dynamically
active electrons in the normal state at T$_c$: 
all dynamically active carriers at T$_c$ are finally subsumed 
into the condensate at zero temperature.

\section{Acknowledgments}
Communications with V. Taufour, N. Curro, D. N. Basov, C. Homes, and D. B. Tanner
are gratefully acknowledged. An extensive set of slides was provided by
D. B. Tanner, containing much unpublished data describing their data
and analysis. K. Koepernik assisted with technical
information. 
%I acknowledge with great appreciation a series of lectures provided
%by R. Kubo at Stony Brook University ca. 1974-1975 with substantial
%equations about rigorous linear response, presented with obvious enthusiasm, which 
%I have not entirely forgotten.


\begin{thebibliography}{27}
\bibitem{explainNb}
The Nb value is the
r.m.s. value of three experimental values, which have a spread of
3 eV. Experimental values can be affected by sample and surface conditions,
and to different frequency ranges of fitting the dielectric function,
often none being in the far-IR at that time hence likely having some interband
contribution.
\end{thebibliography}
\end{document}